# *Mexican Computers: A Brief Technical and Historical Overview*[1]

Daniel Ortiz-Arroyo[2]

## ABSTRACT

*The emergence of the microprocessor in the early 1970s allowed the design of computers that did not require the substantial economic resources of large computer companies of that era. Shortly after this event, a variety of computers based on microprocessors appeared in the United States and other developed countries. Unlike in those countries, where small and large companies developed most personal computers, in Mexico, the first microprocessor-based computers were designed within academic institutions. It is little known that Mexican computers of that era included a variety of systems ranging from purpose-specific research and teaching-oriented computers to high-performance personal computers. The goal of this article is to describe in detail some of these Mexican computers designed between the late 1970s and mid-1980s.*

## I. INTRODUCTION

Mexico began its foray into the use of digital computer technology on June 8, 1958, when the National Autonomous University of Mexico (UNAM) acquired an IBM-650 computer. This date marks a milestone in the history of computing in Latin America, as the IBM-650 was the first electronic computer to operate on this continent, south of the Rio Grande. We commemorate, in 2008, the 50th anniversary of this significant historical event.

Research and development in informatics in Mexico began in the late 1970s. Among the Mexican universities that developed early projects related to computer design were UNAM, the National Polytechnic Institute (IPN), and the Autonomous University of Puebla (BUAP).

By the end of the 1970s, low-cost personal computers were having a strong impact worldwide, making the use of this technology accessible to an increasing number of users. This impact would lead them, by the turn of the next decade, to the overwhelming conquest of the computer market, dominated until then by mainframes and minicomputers. This, coupled with the prevailing sense of nationalism in the national policies of that time, led certain sectors within the Mexican government and academia to be interested in developing computer technology as a strategic step to reduce the country's very high technological dependence on foreign sources. However, during the 1980s, the country's economic situation was notably unstable. Periodic devaluations of the Mexican peso were followed by hyperinflationary processes that led the country into a severe economic crisis. As a result, by the end of 1989, the peso had devalued to about one-hundredth of its initial value at the end of 1981 against the US dollar [1]. The technological development situation in the country was further complicated because investment in science and development was, then as now, well below global standards. Moreover, private investment in these areas was virtually nonexistent. The vast majority of Mexican companies imported all the computer technology they needed [2, 3, 4]. In contrast, a small group of researchers operating from the academic sector had the knowledge and the necessary economic and human resources to develop Mexican computer technology [5, 6].

---

[1] This paper is a translation of the original paper in Spanish "Computadoras Mexicanas: Una Breve Reseña Técnica e Histórica" by Daniel Ortiz Arroyo, Francisco Rodríguez Henríquez, Carlos A. Coello Coello

[2] The author is with the Department of Energy at Aalborg University, Denmark

Thus, during the seventies and eighties, Mexico had a historic and - in light of the information we have now - irreplicable opportunity to join the select group of countries manufacturing computers with their technology. This happened just as this industry was taking off towards an exponential development destined forever to change the lifestyle of the inhabitants of our planet. Despite some successes by the Mexican government, which were met with embryonic developments of prototypes in the Mexican academic sector, we can affirm, from the perspective provided by almost 30 years of history, that Mexico could not seize such an opportunity. Various factors contributed to the failure of the Mexican industry and academia in developing a national computer industry capable of creating its own technology.

The main purpose of this article is to present a brief technical and historical overview of the development of computing in Mexico. Given the breadth of the topic, we will focus primarily on describing Mexican computers designed in the period between the late 1970s and mid-1980s. The creation and design of these computers indicate that, despite adverse economic circumstances prevailing in the country at that time, an incipient high-level computer technology was being developed within the Mexican academic sector. As described in the rest of this article, Mexican computers of that era include a small but rich (and sometimes astonishing) variety of systems ranging from research and teaching-oriented computers to high-performance personal computers.

The computers described in the following sections were selected according to the following criteria: 1) sufficient information existed to allow us to describe their architectures in some detail, and 2) the characteristics of these computers included an original design.

The article is organized as follows. In Section II, we present a summary of the development of computing as a discipline within Mexico. Next, in Section III, we describe in detail the Mexican computers developed in the reference period. Finally, in Section IV, we present our conclusions.

## II. COMPUTING IN MEXICO: THE EARLY YEARS

In the mid-1950s, a group of scientists from UNAM sent engineer Sergio Beltrán López to visit the main campus of the University of California, Los Angeles (UCLA). The main purpose of this historic trip was to understand how IBM-manufactured computers efficiently solved a complex system of simultaneous integral-differential equations. This research problem caught the attention of the UNAM scientific team due to its direct application to solving a significant number of real problems in soil mechanics afflicting Mexico City. Without the help of computers, it was considered that solving these problems would take an unacceptably long time.

As a result of this historic journey, Engineer Beltrán López became convinced of the value of digital computers in solving serious research problems. However, it was not until after intense lobbying that Beltrán López could finally dispel the initial doubts of the UNAM academic community about the use of this new technology. Along with Engineer Beltrán López, Doctors Carlos Graeff

Fernández and Alberto Barajas Celis, professors at that time in the Faculty of Sciences at UNAM, strongly supported the project on the use of computers in scientific research.

The initial plan of the group was to buy an IBM-704 computer equipped with the latest technology. However, even after obtaining a special discount of over 60% from IBM, the price of that computer was well above the budget allocated by UNAM authorities for the project. For this reason, there was no choice but to buy the previous model, an IBM-650[3], which had appeared in the US market in 1954[4].

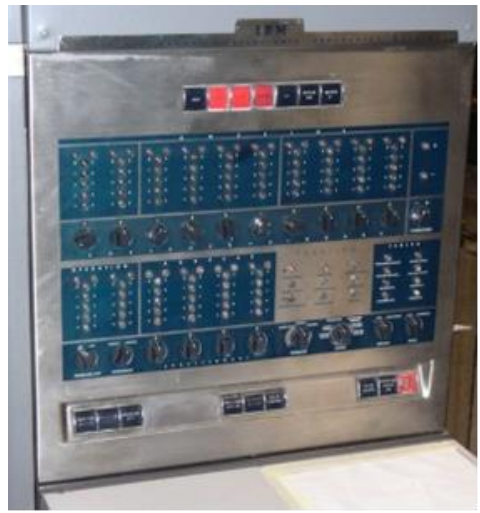

**Figure 1:** Front panel of the IBM-650 computer

We can say then that on June 8, 1958, the official history of computing in Mexico (and Latin America in general) began when UNAM put the IBM-650 into operation. The computer was placed under the care of the Electronic Computing Center (CCE), located in the basement of the old Faculty of Sciences. Its first director was Engineer Beltrán López, and among its collaborators were Renato Iturriaga, Manuel Alvarez, Lian Karp, Javier Treviño, Luis Varela, and Eduardo

---

[3] The IBM-650 acquired by UNAM was a 2nd hand computer whose first owner was UCLA.

[4] In 1958, there were 2,000 IBM-650 computers operating worldwide.

Molina. The IBM-650 operated with vacuum tubes, using a magnetic drum with a capacity of 20,000 digits. It could perform 1,300 addition and subtraction operations per second and operated with a card reader and punch, adopting a numerical system called bi-quinario (see Figure 1). It used an assembler called SOAP (Symbolic Optimizer and Assembly Program), a pseudo-compiler called RUNCIBLE, and an interpreter called BELL [7, 8]. The first tasks assigned to this computer included solving problems in astronomy, physics, and chemical engineering. A database was even created for an anthropology group[5].

Shortly after its creation, the CCE began disseminating knowledge about the applications of the new computer technology. Thus, an annual conference called "Computers and Their Applications" was organized. It is interesting to note that the third edition of that conference, held in 1961, featured lectures by professors John McCarthy, Marvin L. Minsky[6], and Harold V. McIntosh [9].

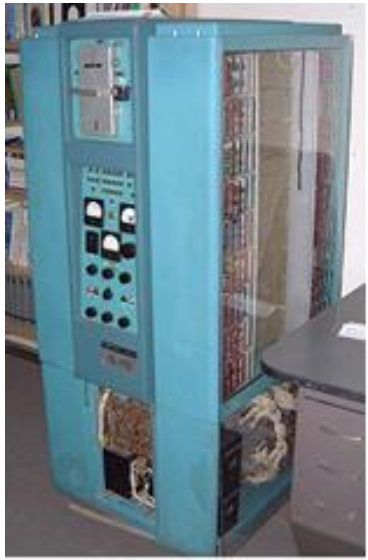

**Figure 2**: A Bendix G-15 computer

In the following years, UNAM purchased other more sophisticated computers [9,

---

[5] It is important to mention that in the same year the Autonomous University of Nuevo Leaon bought another IBM-650 [8].

[6] McCarthy y Minsky are some of the founders of Artificial Intelligence.

10, 11]. For example, in the late 1960s, UNAM acquired a Bendix G-15 (see Figure 2). Part of the design of the Bendix G-15 model contained transistors, had a magnetic tape unit for data storage, a punched card reader, as well as a console for entering programs [7]. This computer was part of the "Mobile Computing Center" educational program, whose main purpose was to disseminate computer knowledge throughout the country.

Other academic institutions such as IPN and the Technological Institute of Higher Studies of Monterrey (ITESM) soon joined the select group of computer users by acquiring an IBM-709 and an IBM-1620, respectively [12, 13][7]. Similarly, other government institutions such as the Mexican Social Security Institute (IMSS), the Federal Congress, the Federal Electricity Company (CFE), and Petróleos Mexicanos (PEMEX), among others, were pioneering Mexican institutions in the digital age.

It has been estimated that by 1968, about 200 computers were operating in the country [8][8]. This quantity is slightly higher than the number of computers that existed at that time in other Latin American countries such as Argentina and Chile [14, 15].

Some years after Mexico entered the digital revolution, several universities decided to offer undergraduate and graduate programs in computer engineering and science. Apparently, the first computer engineering undergraduate program was offered from 1965 by IPN [12, 17]. Shortly thereafter, other institutions such as ITESM (in 1968) [18], the Autonomous University of Puebla (UAP), and the Autonomous University of Nuevo León (UANL) (en 1973) would start their own study programs [17, 19].

At the postgraduate level, it was at UNAM that some computer science courses began to be taught during the early years of the development of this discipline in Mexico. An example of this was the courses taught by Dr. Alejandro Medina

---

[7] In 1958 la UNAM could not acquire the IBM-709 model because of its high cost, the same computer was donated six years after, in 1964, by IBM to the Centro Nacional de Cómputo (CENAC) del IPN.

[8] Apparently less than 18 computers were operating within universities and Mexican institutions of higher education [13].

Plascencia to physics and mathematics students. However, it should be noted that these courses did not officially belong to a formal computer science master's program [20]. Formally, the first Mexican postgraduate program in computer science was founded at UNAM in the early 1960s, and sponsored by UNESCO. In this initial master's program, students from the School of Sciences and Engineering at UNAM were mainly enrolled [8, 20]. By the year 1970, not only UNAM but also Chapingo University and Iberoamerican University would begin to offer a master's program in computer engineering [2, 20, 21].

Traditionally, in Mexico, the majority of serious research work has been carried out within the public sector, either in existing research centers or in government institutions such as CFE or PEMEX. Notably, UNAM, IPN, UAP, CINVESTAV-IPN, and INAOE stand out among the first institutions that conducted research in areas related to computer engineering and computer science. According to the data available to the authors, apparently, the first internationally published Mexican article in computer-related areas was written by CFE engineer Raúl Pavón in 1958 under the title: "The Mexican Light and Power Company Introduces a Direct Way for Fast Computation of Industrial services with power factor adjustment." This article describes an alternative numerical method for the fast computation of the square root of a number used in calculating the power factor in transformers [22].

Similarly, the article published in 1966 by Adolfo Guzmán Arenas and Harold V. McIntosh about CONVERT (a LISP-based language designed by them) seems to be the first Mexican publication to appear in an international journal [23] in computer-related areas.

**III. MEXICAN COMPUTERS**

In Mexico, the development of computer technology began a few years after the appearance of the microprocessor in the United States. What motivated this development was that, from then on, the low cost of microprocessors made it feasible to design a variety of computers. In this section, we describe in detail some of the Mexican computers developed from the late 1970s to the mid-1980s.

For comparison, it is worth noting that during the period covered in this work, various companies in the United States were developing personal computers based on microprocessors, such as the IBM-PC (1981), IBM-XT (1983), and IBM-AT (1984) by IBM, and the Apple III (1980) and Macintosh (1984) by Apple Computer, to name just a few examples of the most popular computers of the time.

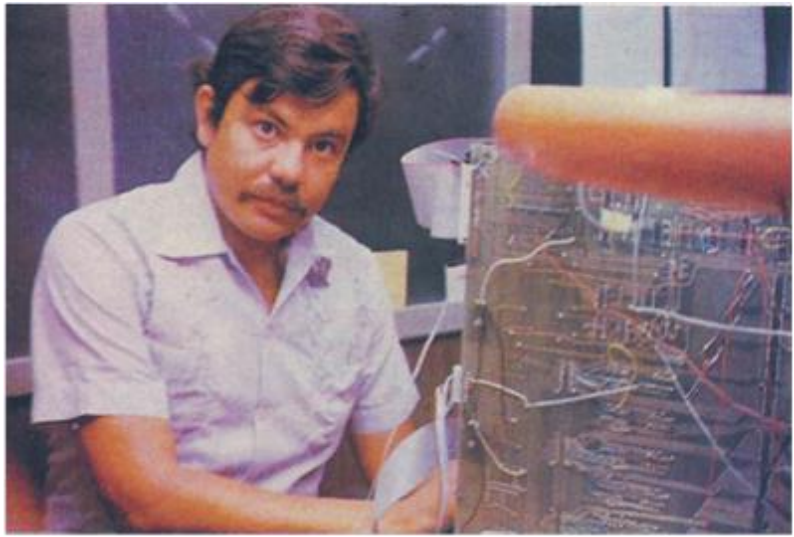

**Figure 3**: Adolfo Guzmán Arenas with the AHR computer]

**III-A Hierarchical Parallel Processing Computer**

One of the first computers designed in Mexico was the Heterarchical Parallel Processing Computer (AHR) [24-28][9], which was built at UNAM in the period 1979-1982. The project was led by Adolfo Guzmán Arenas.

The AHR was designed specifically to efficiently execute programs written in LISP. This computer could accommodate from 5 to 64 Z-80 processors working simultaneously in the execution of a program written in the LISP language (see Figure 3).

The AHR lacked an input/output system as it was used as a slave processor to a minicomputer. For this reason, an operating system for the AHR was never developed. The minicomputer with its operating system was used to edit, load, and observe the output of a program when executed by the AHR [27-28].

---

[9] The term Heterarchical was introduced to indicate that the processors in the architecture were organized in a non-hierarchical form.

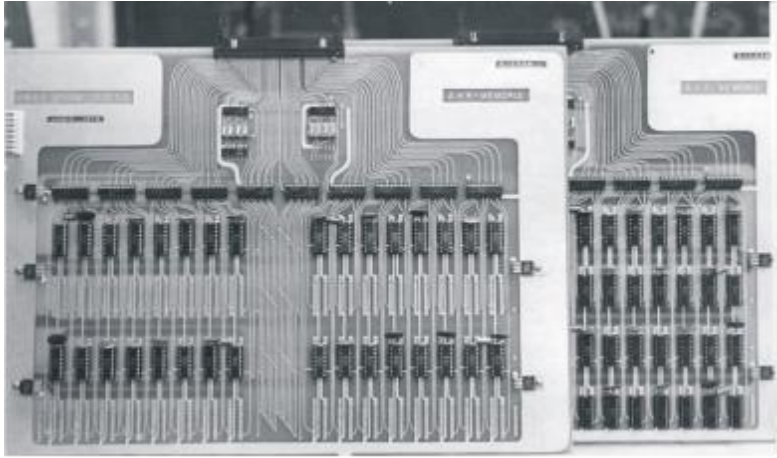

**Figure 4**: View of the FIFO and the AHR grid

The AHR was designed to exploit the parallelism implicit in programs written in the LISP programming language. The multiple processors in the AHR performed parallel evaluation of the arguments of functions written in LISP. Because the arguments of a function in LISP can be other functions or variables, it is possible to generate a representation in the form of a data flow graph of a program in LISP. The nodes in the graph represented functions or variables, and the arcs represented dependencies between nodes. The nodes contained a variable used as a global counter that served to indicate when the arguments of a function in LISP were "ready," an event that allowed the function to be evaluated. The arguments of a function were marked as "ready" when the counter indicated that either of the following two conditions had been met: 1) the variables used as arguments already had an assigned value and/or 2) the functions used as arguments themselves had all their arguments ready for evaluation. When these conditions were met, the node representing a function in the data flow graph was marked as "ready" to be executed.

The nodes in the data flow graph were placed in a shared RAM, called "active memory." Another special RAM bank called "passive memory" held the primitive data types of a program in LISP, i.e., atoms and lists, in addition to the temporary results generated during program execution. The AHR could parallelize the execution of a program written in LISP as follows: Initially, the program was loaded into "passive memory." Then, the data flow graph was constructed in memory. The nodes representing LISP expressions in the program were loaded and copied into "active memory."

A hardware structure of the "First In, First Out" (FIFO) type contained pointers to the nodes residing in "active memory" (see Figure 4). Nodes ready for execution remained at the head of the FIFO, while new nodes dynamically changing state as marked "ready" were placed at the tail of the FIFO.

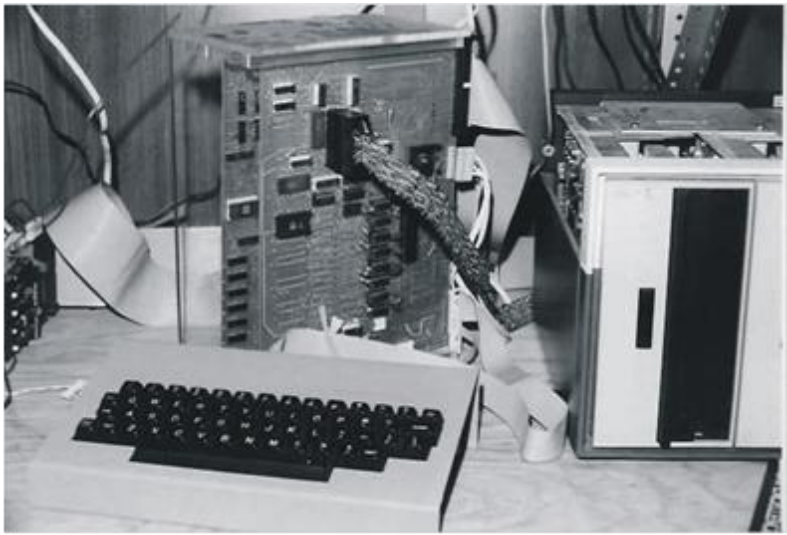

**Figure 5**: One of the LISP processors of the AHR

Another special hardware structure called the "distributor" managed the FIFO. The "distributor's" function was to take nodes that were "ready" from the head of the FIFO and assign them for execution to those Z80 processors (referred to as "LISP processors" in the original article [24-26]) that were free or in a waiting state. To perform this task, the "distributor" used an arbitration circuit that selected a processor within the group of processors that was free, using a static priority system for this purpose. The "distributor" assigned higher priorities to processors physically closer to it. It's worth noting that processors were physically arranged in a ring.

The "LISP processors" (see Figure 5) had shared access to "passive" and "active" memories. These processors informed the "distributor" when they could evaluate new functions. When a "LISP processor" indicated it was ready, it simultaneously sent the result of a previous evaluation to the "distributor." This was done to make communication between processors and the "distributor" more efficient. The results of functions executed by the "LISP processors" were collected by the "distributor" and stored in nodes residing in "active memory" that required them, simultaneously updating the global "ready" argument counter.

The AHR used two data buses to communicate with the "distributor" and "LISP processors." A high-speed bus was used to load into the private memory of "LISP processors" an expression or LISP function ready for evaluation. The same bus was also used to send the result of the previous evaluation of a function executed by a "LISP processor." On the other hand, the low-speed bus was used to send to all "LISP processors" the identification of a program that needed to be stopped or aborted.

The software developed for the AHR consisted of developing an interpreter for the LISP language, written in PL2 and then compiled into Z80 code. The interpreter was loaded into the private memory of each "LISP processor." Finally, the garbage collection mechanism, which is part of most LISP implementations, was executed by a special program on the master minicomputer.

The first prototype of the AHR was completed in late 1982 as a proof of concept for the machine. Unfortunately, plans to continue the further development of the computer were canceled shortly afterward. The data available to us indicates that the AHR was the first research project in the design of digital computers carried out within Mexico.

## III-B. Computers Developed in the Department of Microcomputer Applications at UAP

The Department of Microcomputer Applications at the Institute of Sciences at UAP was one of the pioneering centers in the development of microprocessor-based computers in Mexico[10]. Under the leadership of Harold V. McIntosh, this research center designed and built various computing systems, in addition to compilers and several scientific application programs.

### III-B.1 Multi-User System

The Multi-User System (SMU) consisted of an S-100 bus to which several cards designed in the department were added, such as a floating-point arithmetic coprocessor and interfaces for floppy disk units. Additionally, SMU allowed the connection of multiple Televideo terminals and printers. To achieve this, the CP/M operating system was disassembled to increase its functionality, enabling it to handle multiple users [17,29]. Similarly, the Fortran language compiler, called Fortran 80, was disassembled to modify and add code to take advantage of the developed arithmetic coprocessor. The SMU was developed between 1979 and 1983 (see Figures 6 and 7).

---

[10] It is important to highlight that the computer development projects carried out within the Microcomputer Application Department of the BUAP had special support from Luis Rivera Terrazas, rector at the time of the BUAP and pioneer of astronomy in Mexico.

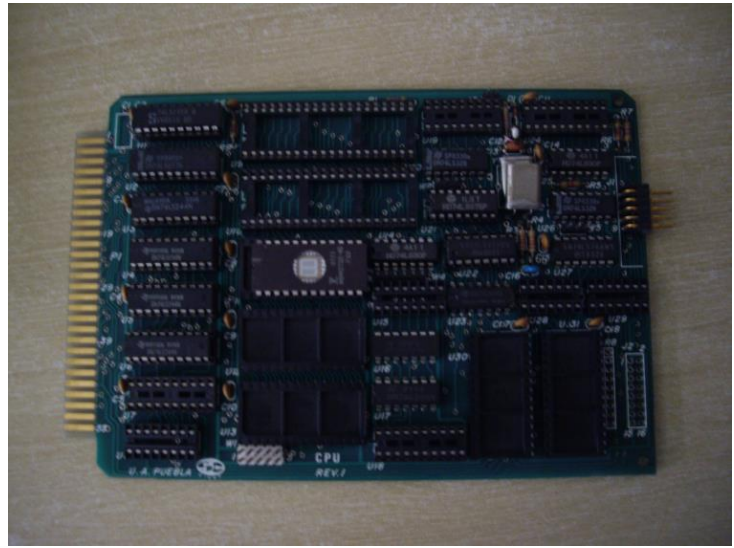

**Figure 6**: SMU Motherboard

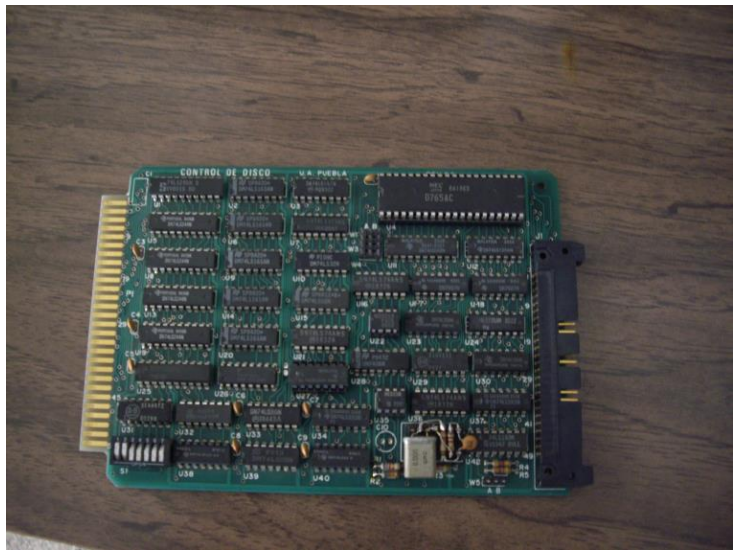

**Figure 7:** SMU Disk Control Card

**III-B.2 CP-UAP System**

The personal computer known as CP-UAP consisted of the design of a motherboard using an NEC-V20 processor on an STD bus. The NEC-V20 processor, manufactured by NEC, was compatible with Intel's 8088 and 80186 processors in terms of physical interface and instruction set, respectively.

Additionally, the NEC-V20 could operate at a clock frequency up to twice as high as the Z-80 microprocessor. The project included the design of memory cards and a video interface for the computer. The CP-UAP was developed between 1984 and 1986 and used the CP/M operating system [29].

**III-B.3 Turing-850 Personal Computer**

The Turing-850 personal computer (see Figure 8) was specially designed with technology that could be easily produced within Mexico. The original idea of the project was to transfer the developed technology to Mexican companies interested in mass production. The project began in mid-1981, and the first prototype of the computer was completed in late 1984 [30].

The design of the Turing-850 included some novel features for that time, such as a hierarchical architecture based on two Z-80 processors manufactured by Zilog Inc. and a light pen that allowed drawing on the screen. The Z80 processors were chosen for their compatibility with the CP/M operating system, the only personal computer operating system available in the country at that time.

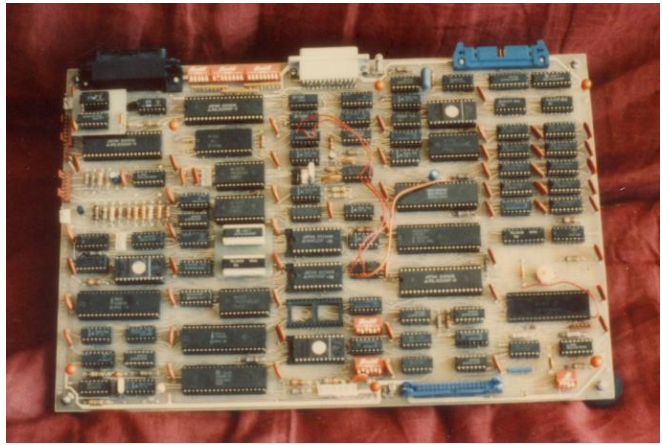

**Figure 8**: Turing-850 Main Board

The processors in the Turing-850 were interconnected in a master-slave configuration (see Figure 8). The goal of this architecture was to produce a high-performance computer by dividing the processing of input/output instructions between the two processors. Some of the fastest processors available on the market at that time were used in the design. The idea of using two processors was to give the Turing-850 a competitive advantage, in terms of performance, compared to other personal computers available on the market. Similarly, due to the increasing popularity of computer networks at that time, the slave processor was planned to handle network communication functions in subsequent models of the Turing-850.

The master processor (also called the central processor (CP)) in the Turing-850 was a 6 MHz Z80A responsible for executing the operating system, user programs, and handling floppy disks. The design also included a Direct Memory Access (DMA) channel used to allow secondary memory devices direct access to memory without requiring intervention from the central processor. The main memory system consisted of 64KB of dynamic RAM.

A 4MHz Z-80B, called the peripheral processor, managed the rest of the input/output devices. This processor-controlled serial and parallel ports, along with the video terminal. Coordination between the central processor and the peripheral was established through a parallel port using a special communication protocol. The central processor obtained instructions from memory, determined which operations should be executed locally, and which should be sent to the peripheral processor, which handled receiving the request, executing the indicated input/output operation, and sending the result to the central processor. The peripheral processor used 24KB of static RAM to store its internal data.

The design of the Turing-850 also allowed the connection of other peripheral devices through a special external connector that provided direct access to the peripheral processor's bus.

The software developed for the Turing-850 was written directly in Z80 assembly language and allowed loading and running the CP/M operating system along with all available applications for it.

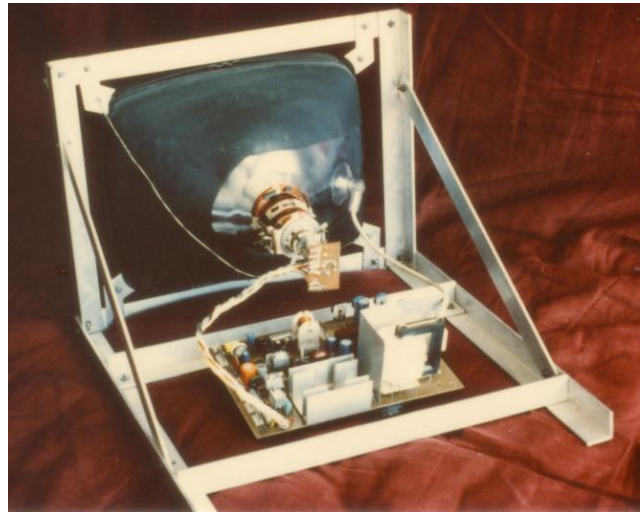

**Figure 9:** Turing-850 Monitor

The video monitor of the Turing-850 used a cathode-ray tube (CRT) manufactured by one of the local television companies (see Figure 9). To control the CRT, special amplifiers were designed since the computer monitor required a higher bandwidth than that used by normal televisions.

To avoid the use of expensive external voltage regulators that were commonly used at that time to protect all electronic equipment from sudden voltage variations, a self-regulated power source was designed (see Figure 10). The source contained a standard linear regulator, along with a feedback circuit that controlled the amount of power delivered to the source using a thyristor.

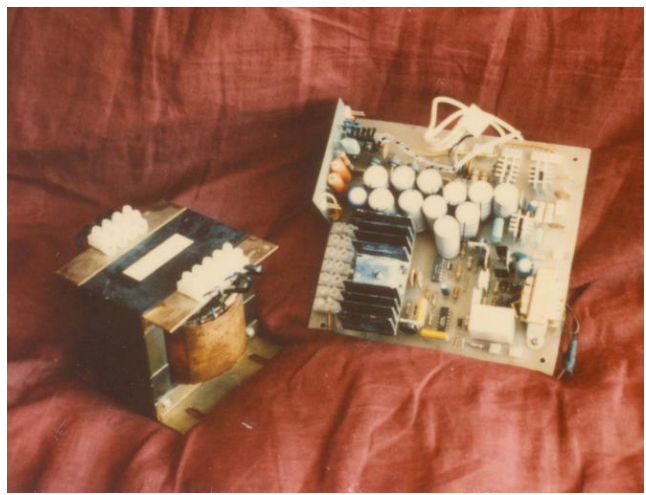

**Figure 10:** Turing-850 Power Supply

The Turing-850 cabinet was designed following the ergonomic standards of the time. The cabinet, constructed with metal sheets, housed all components except for the keyboard. The final design of the Turing-850 (see Figure 11) had a content of national components close to 65%, allowing the machine to be produced within the country without difficulties.

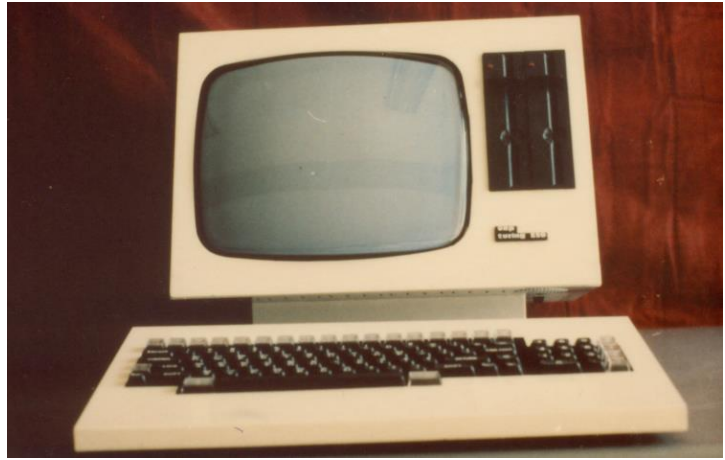

**Figure 11:** Front View of Turing-850 Final Prototype

Unfortunately, after completing the first prototype of the Turing-850, and despite presenting the project at various academic and industrial forums, it was impossible to convince any Mexican company to mass-produce the computer. The original plans of the project were to continue designing a 32-bit computer for which initial specifications had already been written [31]. However, due to the ongoing economic instability of the country at that time, this project was prematurely canceled.

The Turing-850 project was led by Luis Medina-Vaillard, and those who participated in its development included Gregorio Arenas Muñoz, Carlos Blanco Salinas, Sergio Guevara Rubalcava, Daniel Ortiz Arroyo, and Francisco Serrano Osorio.

**III-C The IPN E-16 Computer**

In August 1984, the first prototype of the Almita II computer was completed at the Technological Research Center in Computing of the National Polytechnic Institute (CINTEC-IPN), with Dr. Miguel Lindig Bos as its principal designer. In the 1984 version, Almita II had 256 KB of RAM, two 5.25-inch floppy disk drives with a capacity of 360 KB each, and an intelligent terminal with an Intel 8031 processor. The central processor of this computer was a 16-bit Intel 80186 operating at a speed of 8 MHz. At that time, this processor represented the state of the art in technology. The design of Almita II was a notable achievement in Mexican engineering, experimentally proving that its processing speed was up to 3.4 times higher than that of IBM's first personal computer launched on the market [30, 31].

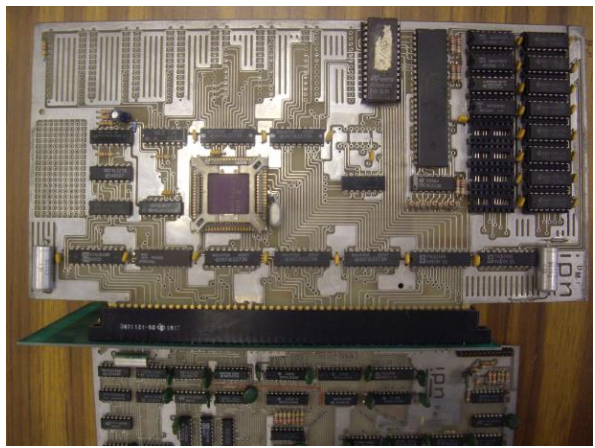

**Figure 12:** Mother board of IPN-E16

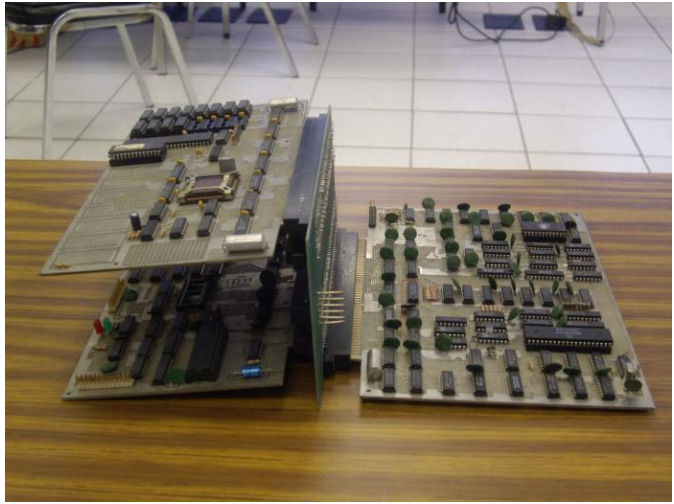

**Figure 13:** Side view of IPN E-16

It is worth mentioning that the 1984 prototype of the Almita II computer was more proof of concept. The computer did not have its own casing, and the different computer cards were not assembled using printed circuits but rather wire-wrap techniques. Despite intense negotiations, it was not until July 1986 that Dr. Lindig Bos received approval from the authorities of the National Polytechnic Institute to build 10 personal computers. This task was successfully completed at the end of 1986 when the computers were exhibited as part of the commemorative events for the fiftieth anniversary of the National Polytechnic Institute held from November 26 to December 10, 1986.

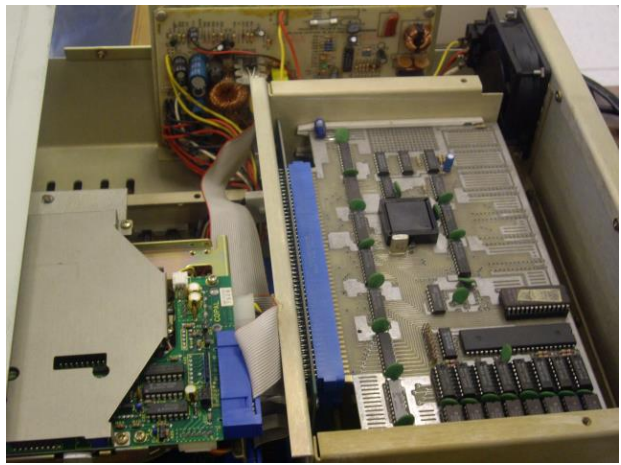

**Figure 14:** Inside view of IPN E-16 Cabinet

The computer model presented in the November-December 1986 exhibition had significant differences from the original 1984 design. The most significant change was the selection of the Intel 80188 integrated circuit for the 1986 model. This choice allowed for a more compact design since the Intel 80188 includes various peripheral components that enable it to operate almost autonomously. However, the selection of this processor implied a slight loss in the computer's performance because the internal bus of the 80188 has an 8-bit word size. Nevertheless, the computer presented in the 1987 exhibition had a speed 2.4 times higher than the IBM-PC personal computer [33]. Additionally, the new model had 16 memories of 256 Kbits each, representing a RAM space of 512 KB. Due in part to the changes made in the architecture of Almita II, it was decided to rename it as the IPN E-16 personal computer.

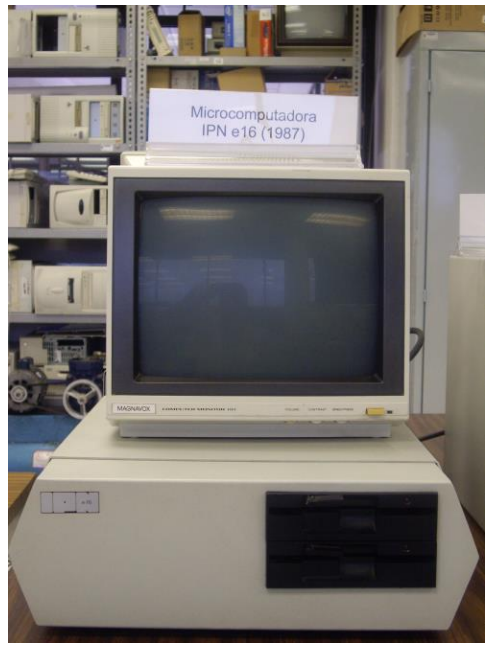

**Figure 15:** Front view of the final prototype of IPN E-16

The introduction of the IPN E-16 in 1987 had a profound impact on the institute's authorities, to the point that they decided to start mass production of this computer to self-supply the information technology needs of middle and higher education schools at the National Polytechnic Institute. The program was so successful that by the end of 1993, more than 1,189 IPN E-16 computers and its descendants were in daily operation in the majority of the National Polytechnic Institute's departments, providing support for administrative and teaching activities [34, 36]. Figure 12 shows the motherboard of the IPN E-16, while Figure 13 displays the set of IPN E-16 cards, including the motherboard, the controller card for devices, and serial and parallel ports. Figure 14 provides an interior view of the IPN E-16 cabinet, and finally, Figure 15 presents a front view of the final prototype of the IPN E-16.

It is not an exaggeration to state that the IPN E-16 computer constitutes a rare partially success story in the history of Mexican digital designs.

**Table 1.** Summary of the characteristics of the discussed Mexican computers in this article

|  | **AHR** | **SMU-BUAP** | **Turing-850** | **IPN E-16** |
|---|---|---|---|---|
| **Operating System** | None | CP/M | CP/M | MS-DOS |
| **Software written** | LISP intepreter | Modifications to CP/M and a Fortran compiler | BIOS written in Z80 assembly language | Generic |
| **CPU** | Z80@6MHz | NEC-V20 @8MHz | Z-80@6MHz | 80188@8MHz |
| **# de processors** | Up to 64 | 1 | 2 | 1 |
| **RAM** | 256KB | 256KB | 256KB | 512KB |
| **# users** | 1 | 10 | 1 | 1 |
| **Case** | None | Metal | Metal | Metal |
| **Computers build** | 1 | <5 | 1 | 1189 |
| **Period of development** | 1979-1982 | 1979-1983 | 1982-1984 | 1984-1987 |

**IV. Conclusions**

The widespread distribution and low cost of microprocessors in the 1970s allowed for the design of a wide variety of computer systems. This fact was especially significant for developing countries like Mexico, which did not have the economic and human resources of large computer companies.

In this work, we have described some of the computers developed in Mexico from the late 1970s to the mid-1980s. These computers are examples of the level of technological development that was taking place in Mexico in the field. The described Mexican computers show that, despite the adverse economic conditions prevailing at that time, it was feasible to develop state-of-the-art computer technology within public universities.

Unfortunately, the crisis of the 1980s, the rapid advance of technology, the lack of collaboration and coordination among the various groups designing computers in Mexico, and the lack of communication between industry and academia hindered the country's further development in this field.

In contrast to Mexico, Brazil, Singapore, South Korea, and Taiwan invested heavily in technological development in the 1980s. Brazil's case is significant because this country has many similarities to Mexico in terms of its economic and industrial development. Unlike Mexico, Brazil had relative success in creating a computer industry capable of producing its own technology. This was partly due to the decisive support from the public and private sectors to create and continue developing a national computer industry.

**ACKNOWLEDGMENT**

The author would like to acknowledge the contribution of Francisco Rodríguez Henríquez and Carlos A. Coello Coello in writing parts of the first version of this paper in Spanish. We would also like to thank Dr. Adolfo Guzmán Arenas and the professors Hugo García Monroy, Eduardo Rodríguez Escobar, and Juan Carlos González Robles for being interviewed and sharing their experiences during the development of their respective designs. They also appreciate the permission to publish images of the AHR, SMU-BUAP, and IPN E-16 computers. Thanks, are also extended to Engineer Partida-Romo of the IMP for providing information regarding the design of the Impetrón computer.